\documentclass{aa}

\usepackage{natbib}
\usepackage{epsfig}
\usepackage{txfonts}
\usepackage{longtable} 

\begin{document} 

\newcommand{\chisq}{\mbox{$\chi^2$}}
\newcommand{\es}{erg s$^{-1}$}                          
\newcommand{\ecms}{erg cm$^{-2}$ s$^{-1}$}              
\newcommand{\ks}{ks$^{-1}$}       
\newcommand{\cmtwo}{cm$^{-2}$}
\newcommand{\msun}{M$_{\odot}$} 
\newcommand{\xmm}{XMM-\emph{Newton}} 
\newcommand{\chandra}{\emph{Chandra}} 
\newcommand{\spitzer}{\emph{Spitzer}} 
\newcommand{\nh}{\mbox{$N({\rm H})$}}
\newcommand{\rhoph}{$\rho$~Oph}
\newcommand{\x}{$_{\rm xmm}$}
\authorrunning{G. Giardino et al.}
\titlerunning{The X-ray luminosity of solar-mass stars in NGC 752}

\title{The X-ray luminosity of solar-mass stars in the intermediate age open cluster NGC 752}

\author{G. Giardino\inst{1} \and I. Pillitteri\inst{2} F. Favata\inst{3} \and
  G. Micela\inst{4}}

\institute{ESA -- Research and Science Support
  Department, ESTEC, 
  Postbus 299, NL-2200 AG Noordwijk, The Netherlands
\and
DSFA, Universit\`a degli Studi di Palermo, Piazza del Parlamento 1, 90134, Palermo, Italy
\and
ESA -- Planning and Community Coordination Office, Science Programme, 8-10 rue Mario Nikis,  F-75738 Paris Cedex 15, France
\and
INAF -- Osservatorio Astronomico di Palermo, 
Piazza del Parlamento 1, I-90134 Palermo, Italy
}

\offprints{G. Giardino}

\date{Received date / Accepted date}

\abstract 
    {}{While observational evidence shows that most of the decline in
      a star's X-ray activity occurs between the age of the Hyades
      ($\sim 8 \times 10^8$ yrs) and that of the Sun, very little is
      known about the evolution of stellar activity between these
      ages. To gain information on the typical level of coronal
      activity at a star's intermediate age, we studied the X-ray
      emission from stars in the 1.9 Gyr old open cluster NGC~752.}
    {We analysed a $\sim$ 140 ks \chandra\ observation of NGC~752 and
      a $\sim$ 50 ks \xmm\ observation of the same cluster. We
      detected 262 X-ray sources in the \chandra\ data and  145 sources
      in the \xmm\ observation.  Around 90\% of the catalogued
      cluster members within \chandra's field-of-view are detected in
      the X-ray. The X-ray luminosity of all observed cluster members
      (28 stars) and of  11 cluster member candidates was derived.}
    {Our data indicate that, at an age of 1.9 Gyr, the typical
          X-ray luminosity of the
      cluster members with $M=0.8-1.2~M_{\sun}$  is $L_{\rm X}
        = 1.3 \times 10^{28}$ \es, so approximately a factor of 6 less
      intense than that observed in the younger Hyades. Given that $L_{\rm X}$ is proportional to the square of a
      star's rotational rate, the median $L_{\rm X}$ of NGC~752 is
      consistent, for $t \ga 1$ Gyr, to a decaying rate in rotational
      velocities $v_{\rm rot} \propto t^{-\alpha}$ with $\alpha \sim
      0.75$, steeper
      than the Skumanich relation ($\alpha \simeq 0.5$) and significantly steeper than observed between the Pleiades and the
      Hyades (where $\alpha <0.3$), suggesting that a change in the rotational regimes of
      the stellar interiors is taking place at $t\sim 1$ Gyr.}{}

\maketitle

\section{Introduction}
\label{sec:intro}

Investigations of solar-type stars in the better-studied nearby open
clusters have provided a basis for understanding the evolution of
stellar activity and  rotation velocity.
The early {\it Einstein} and {\sc Rosat} studies of open clusters revealed
that the mean X-ray luminosity of G, K and early M stars decreases
steadily with age from pre-main sequence stars (PMS) through the
Pleiades ($\sim 8\times 10^7$ yrs), the Hyades ($\sim 8\times 10^8$
yrs) and the old disc population (\citealp{randich2000};
\citealp{micela2002}).  

In main-sequence stars, late-type stellar magnetic activity is
  regulated principally by rotation (\citealp{pgr+81};
  \citealp{bds+95}) and its decay with stellar age is attributed to
  rotational spin-down. Magnetic braking models indicate that surface
  rotational velocities should decline as $v_{\rm rot} \propto
  t^{-\alpha}$ with $0.38 < \alpha < 0.75$, depending on whether the
  magnetic field geometry is radial or closer to a dipole 
  configuration (\citealp{kawaler88}). Models considering
  possible differences in the pre-main sequence disc-locking time,
  solid-body versus differential rotation in the interior, and the
  onset of magnetic saturation give decay laws over the range $0.2 <
  \alpha< 0.8$ for 1$-$5 Gyr old solar-mass stars
  (\citealp{kpb+97}).

Although \cite{skuma72} found the decay in rotational velocities
  and the strength of the magnetic activity indicators Ca\,{\sc ii} to
  be consistent with pure magnetic breaking ($\alpha \simeq 0.5$) from
  the age of the Pleiades to that of the Sun, more recent evidence
  points to a decay rate of the stellar activity which is less steep
  between the Pleiades and the Hyades and drops significantly faster
  from the Hyades ages to that of the Sun or field stars
  (e.g. \citealp{shb85} for the evolution of transition-region lines;
  \citealp{micela02} and \citealp{pf05} for the evolution of coronal
  emission). This may indicate a change, between 1 and 5 Gyr, in the
  rotational regime of stellar interiors.

Helioseismology has shown that in the Sun there is no
difference between the angular rotation of the radiative core and that
of the convective envelope, while models suggest that a significant
radial gradient is present at earlier stages (\citealp{mb91},
\citealp{bcm99}, \citealp{silpin00}).  \cite{qam+98} reported the
  surface rotational evolution between the Pleiades and the Hyades to be less
  steep than the \cite{skuma72} relation, and interpreted
  this
  as evidence of angular momentum transport from the core to the
  envelope.
Determination of the evolution of coronal emission beyond
the Hyades age can therefore be an effective probe of the evolution of
stellar interiors in older stars.

 Although  most of the decline in X-ray activity occurs between the age
of the Hyades and that of the Sun, the evolution of stellar activity in
solar-type stars older than the Hyades is still poorly known.  For stars
older than the Hyades, the data on coronal emission level are based on
field stars and the Sun.  While the Sun's age is well constrained at
4.5 Gyr, it is only one star and therefore cannot be assumed
to be representative of the behavior of solar-type stars at its
current age. The age of field stars are poorly constrained and they
represent in general a mixture of stellar populations and chemical
composition.

The best approach to explore the stellar activity of solar-type stars
at an age intermediate between those of the Hyades and the Sun is the
X-ray observation of an old open cluster, which naturally provides a
sample of similar aged stars, at the same distance and with similar
chemical abundances. This obvious approach is difficult for
observational reasons: there are few old clusters in the solar
neighbourhood, since many of them are dissipated in few billion years
through dynamical evolution, and their intrinsically low X-ray
luminosity requires extremely long exposures with current X-ray
instrumentation.  The only known, sufficiently nearby, intermediate
age, open cluster is NGC~752. At a distance of $\sim$ 400 pc and with
age between 1.7 and 2 Gyr, this is the only open cluster in which one
can measure the X-ray luminosity of individual dG stars with exposure
times below 200 ks -- unfortunately, though, the cluster is rather
dispersed.

We present here the results of the analysis of two X-ray observations
of NGC~752: a $\sim$ 140 ks \chandra\ observation and a $\sim$ 50 ks
\xmm\ observation.  The analysis focuses on solar-type stars,
  with mass $0.8 < M < 1.2~M_{\sun}$. The paper is organised
as follows. After a summary, below, of the properties of NGC~752, the
observations and data analysis are described in
Sect.\,\ref{sec:obs}. Results are presented in Sect.\,\ref{sec:res}
and the implications are discussed in Sect.\,\ref{sec:disc}.

\subsection{NGC~752}
\label{sec:ngc752}

The most comprehensive study of the open cluster NGC~752 to date is
the one by \cite{dlm+94} (hereafter DLM94). These authors collected
and systematised a number of existing proper-motion and
radial-velocity studies of this cluster, supplemented them with their
own measures of radial velocities, and combined them with existing
photometric and spectroscopic data. Thanks to this effort, they claim
to have obtained an effectively complete census of all probable and
possible cluster members down to the observable limit of the unevolved
main sequence which, in their case, corresponds to $V \sim
13.5-14$. The electronic version of their catalogue available from
{\sc Vizier} lists 255 stars of which 157 are classified as probable
or possible members of NGC~752.

Combining the spectroscopic and photometric approaches, DLM94
establish the cluster metallicity to be [Fe/H] $= -0.15 \pm 0.05$
dex. They also derive a reddening value of $E(B-V) = 0.035 \pm 0.005$,
which corresponds to an absorbing column density of $N({\rm H}) = 2 \times
10^{20}$ \cmtwo\ or $A_V = 0.1$ (\citealp{cox00}).
From intermediate-band photometry and main-sequence fitting, 
DLM94 derive a
distance modulus of $(m-M)=8.25 \pm 0.10$ mag which corresponds to a
distance of $\sim 430$ pc (assuming $A_V = 0.1$). 
Comparing the data with theoretical models, they estimate for this cluster an age of
$1.9\pm0.2$ Gyr from classical isochrones and $1.7\pm0.1$ from
isochrones with over-shooting, in good agreement with the results from
isochrone fitting of \cite{mmm93} (1.8 Gyr) and of \cite{ddg+95} (2.0
Gyr).

Since the first proper-motion study by \cite{ebbi39}, it has been known
that in NGC~752 the unevolved main sequence is deficient in stars
relative to the turn-off. This is likely due to mass-segregation
effects and the cluster relaxation, which leads to the evaporation of
the less massive members.

NGC~752 was observed in the X-ray with {\sc Rosat} by
\cite{bv96}. They catalogued 49 sources, seven of which were
identified with cluster members.

\section{Observations and data analysis}
\label{sec:obs}

\subsection{\chandra\ observation}
\label{sec:cha_obs}

The 135 ks observation of NGC~752 was performed by the \chandra\ ACIS
camera on September 29 2003 starting at 21:11:59 UT. The Principal
Investigator for the observation is T. Simon, who presented
preliminary results in \cite{cs04}. We retrieved the data from the
public archive, with no re-processing done on the archival data.

\begin{figure*}[]
  \begin{center} \leavevmode 

        \epsfig{file=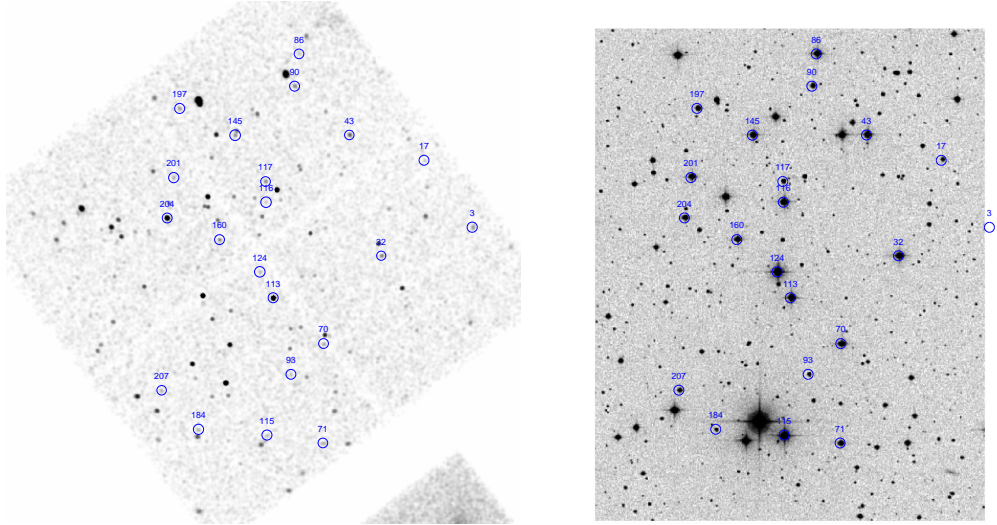, width=14.0cm, bbllx=26, bblly=272, bburx=577, bbury=569, clip=}

\caption{{\it Left} -- \chandra\ image of NGC~752; X-ray sources  which
  are possible or likely members of the cluster are indicated. {\it Right} -- DSS image in the same
direction.}

    \label{fig:image}
  \end{center}
\end{figure*}

We performed the source detection on the event list, using the Wavelet
Transform detection algorithm developed at Palermo Astronomical
Observatory (\textsc{Pwdetect}, available at
http://oapa.astropa.unipa.it/progetti\_ricerca/PWDetect); 
  initially the energy range $0.2-10$ keV was selected and the
  threshold for source detection was taken as to ensure a maximum of
  1--2 spurious sources per field. 169 sources were detected in this
  way. The analysis of these sources hardness ratio showed, however,
  that all the catalogued stars in the field had low hardness ratios, HR
  $\la 0.2$ (where HR is the number of photons in the band $2-8$ keV
  over number in the band $0.5-2$ keV). Thus, to maximise the
  detection of stellar sources, \textsc{Pwdetect} was applied to the
  event list in the energy range $0.5-2$ keV. Using a detection
  threshold which ensures less than 1 spurious source per field leads
  to the detection of 188 sources, while lowering this threshold to 10
  spurious sources per field, allows 262 sources to be identified in
  this energy range. This is a significant increase (well above the
  number expected if all the additional sources were spurious),
  thus we retain this list of 262 sources as our final list of sources
  in the NGC~752 field, with the
  caveat that $\sim$ 10 sources among them are likely spurious. Note that the
    existence of $\sim 10$ spurious sources in the list is not so much of a
    problem in this context, because cluster members or candidate
    members are identified by the existence of a visible or
    near-IR counterpart.

Table~6\footnote{on line.} lists the coordinates
for all  262 detected sources. We searched for 2MASS counterparts
to the X-ray sources using the 2MASS Point Source
  Catalogue (PSC) and a search radius of 3 arcsec and found a
counterpart for 43 sources. Searching within the Point Source Reject Table of
  the 2MASS Extended Mission leads to the further identification of 1
  counterpart (source id 87).  For the sources with a 2MASS
counterpart, the magnitude in the $J$, $H$ and $K$ bands is also
given in Table~6.

We also searched for visible counterparts within the catalogue of
NGC~752 by DLM94 and found  26 counterparts (listed in Table
  \,\ref{tab:vmatches}); of these, 20 are likely members of the
open cluster and 1 is a possible member; the other five are likely
non-members (DLM94). All of the 26 sources with counterparts from
DLM94 also have a 2MASS counterpart.  Indeed, most of the sources with
2MASS counterparts are stars, while most of the sources without a
counterpart are probably extra-galactic sources.  Fig.~\ref{fig:image}
shows the \chandra\ image of NGC~752 towards the cluster centre
(01:57:41 +37:47:06). The catalogued members of NGC~752 are spread over an
area of approximately $90 \times 90$ arcmin while the \chandra\ image
covers a central area of $17 \times 17$ arcmin.

 Finally, we searched for visible counterparts to the X-ray sources
  in the {\sc Nomad} catalogue (\citealp{zml+04}) and found 65 matches
  (within a search radius of 3 arcsec). Of these 65 sources, 42
  sources have also 2MASS counterparts.  All the {\sc Nomad}-matched
  sources without a 2MASS counterpart lack $V$ measurements and have
  $B > 19.5$, apart from one which has $B=18.9$. The use of the {\sc
    Nomad} catalogue, thus, does not allow any additional X-ray source
  with $B \la 19$ to be identified with respect to 2MASS; $B =19$
  corresponds to a star in NGC~752 of less than $0.45~M_{\sun}$
  (according to the isochrones of \citealp{gbb+02}). We expect most of
  the objects with $B > 19.5$ to be extragalactic and, in
  addition, if a few of them were members of NGC~752, their mass
  would be well below the target of this study, which are solar-mass stars
  with $0.8 \la M \la 1.2 M_{\sun}$. Accordingly, we did not further consider
  the {\sc Nomad} catalogue to identify potential members of NGC~752
  among the \chandra\ sources.

\begin{figure}[]
  \begin{center} \leavevmode 

        \epsfig{file=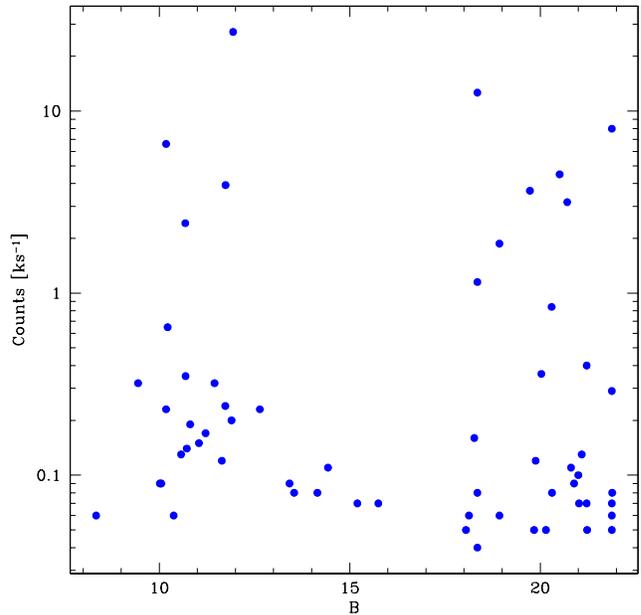, width=9.0cm, clip=}

\caption{Count rate vs. B magnitudes for the \chandra\ sources with a
counterpart in the {\sc Nomad} catalogue.}

    \label{fig:distB}
  \end{center}
\end{figure}

It is interesting, however, to look at the scatter plot of the
  X-ray counts versus $B$ magnitudes for the X-ray sources with a {\sc
    Nomad} counterpart shown in Fig.\,\ref{fig:distB}. A lack of
  sources with $14 < B < 18$ is clearly noticeable and does not appear
  to be due to the sensitivity of the \chandra\ data, (as weaker sources
  are detected for other values of $B$). This indicates that the lack of X-ray
  sources with a counterpart with $14 < B < 18$ is intrinsic to the
  field, consistent with the observation that NGC~752 appears to be
  deficient in low mass stars.

We estimated the number of expected background sources in the
observation using the $\log N-\log S$ relationship for X-ray sources
in \cite{bah+01}.  As in Table\,6, the faintest
source has a typical count rate of  0.035 \ks; assuming for
extragalactic sources a power-law spectrum with spectral index
$\Gamma=1.4$ and a total interstellar absorption column of $N({\rm H})
= 5 \times 10^{20}$ cm$^{-2}$ (\citealp{dl90}; \citealp{kbh+05}), this
count rate corresponds\footnote{Using the {\sc pimms} software at {\sc Heasarc}} to a flux of  2.2~$\times 10^{-16}$
\ecms\ in the
energy band $0.5-2$ keV.  In this energy band and at this flux level,
the expected number density of background sources determined on the
basis of the $\log N-\log S$ relationship given in \cite{bah+01} is
 2100-2610 per square degree. The area imaged by \chandra\ is
0.08 square degree, so that the expected number of serendipitous
extra-galactic sources is between  170 and 210, consistent with the
hypothesis that most of non-identified sources are extragalactic.

For the brightest sources, for which a spectral and timing analysis
was carried out, source and background regions were defined in
\textsc{ds9}, and light curves and spectra were extracted from the
photon list in the energy range 0.3-7.5 keV using CIAO V.~3.3 threads,
which were also used for the generation of the relative response
matrices. The spectral analysis was performed using the \textsc{Xspec}
package V11.2, after rebinning the spectra to a minimum of 5 source
counts per (variable width) spectral bin.

\begin{table}[]
  \begin{center} 

\caption{\chandra\ sources with a visible counterpart from the study by
  DLM94.}
\leavevmode
\begin{tabular}{l|cccc}
\hline\hline
Id & Count rate & DLM94 Id & MC$^{(a)}$ & $r^{(b)}$ \\
~      & ks$^{-1}$ & ~ &  & arcsec \\
\hline
3   & 0.32 $\pm$ 0.09 & 654  & $+$ & 0.49\\
17  & 0.08 $\pm$ 0.03 & 701  & $+$ & 1.61\\
32  & 0.23 $\pm$ 0.04 & 745  & $+$ & 0.32\\
38  & 0.09 $\pm$ 0.04 & 758  & $-$ & 0.75\\
43  & 0.35 $\pm$ 0.06 & 772  & $+$ & 0.50\\
70  & 0.19 $\pm$ 0.07 & 798  & $0$ & 0.25\\
71  & 0.12 $\pm$ 0.05 & 799  & $+$ & 0.67\\
86  & 0.13 $\pm$ 0.05 & 823  & $+$ & 0.46\\
90  & 0.2  $\pm$ 0.06 & 824  & $+$ & 0.25\\
93  & 0.08 $\pm$ 0.03 & 828  & $+$ & 0.70\\
113 & 6.6  $\pm$ 0.24 & 849  & $+$ & 0.03\\
115 & 0.09 $\pm$ 0.04 & 858  & $+$ & 1.84\\
116 & 0.06 $\pm$ 0.03 & 857  & $+$ & 0.26\\
117 & 0.23 $\pm$ 0.07 & 859  & $+$ & 0.09\\
124 & 0.09 $\pm$ 0.03 & 867  & $+$ & 0.46\\
140 & 0.06 $\pm$ 0.03 & 882  & $-$ & 0.41\\
145 & 0.14 $\pm$ 0.05 & 888  & $+$ & 0.03\\
160 & 0.17 $\pm$ 0.05 & 901  & $+$ & 0.21\\
176 & 2.42 $\pm$ 0.14 & 911  & $-$ & 0.23\\
184 & 0.11 $\pm$ 0.05 & 917  & $+$ & 0.46\\
197 & 0.24 $\pm$ 0.07 & 935  & $+$ & 0.89\\
201 & 0.15 $\pm$ 0.05 & 941  & $+$ & 1.30\\
204 & 3.92 $\pm$ 0.21 & 950  & $+$ & 0.12\\
207 & 0.07 $\pm$ 0.03 & 953  & $+$ & 0.29\\
211 & 0.65 $\pm$ 0.08 & 958  & $-$ & 0.08\\
258 & 0.32 $\pm$ 0.06 & 1033 & $-$ & 0.29\\
\hline

\end{tabular}
\label{tab:vmatches}
  \end{center}
$^{(a)}$ star's membership class according to DLM94: $+$
  probable member, $0$ possible member, and $-$ probable
  non-member.

$^{(b)}$ radial distance from counterpart 
\end{table}

\subsection{\xmm\ observation}

NGC~752 was observed for $\sim$ 49 ks by the \xmm\ EPIC camera on
February 5 2003 starting at 23:29:25 UT $-$ nominal pointing 01:57:38
+37:47:60,  thus the \xmm\ field-of-view (FOV) includes the \chandra\ FOV. The
Principal Investigator for the observation is G. Tagliaferri. We
retrieved the ODF data from the \xmm\ data archive and processed them
with SAS 7.1 to obtain tables of photons recorded with EPIC
instruments (MOS1, 2 and PN) calibrated in energy and detection times.
To improve the detection
process for faint sources, we checked the global background light
curve and we filtered out the first part of the observation 
where the background showed rapid variability.  The filtered exposure
times are 40 ks for MOS1 and MOS2, and 39 ks for PN. 

For the source detection, we used the \textsc{Pwxdetect} code
developed at Palermo Observatory and derived from the analogous
Chandra \textsc{Pwdetect} code based on wavelet transform
analysis. This allows the three EPIC exposures to be combined in order to
gain deeper sensitivity with respect to the source detection based on
single images. We detected 145 point sources  in the energy
  band 0.5-2.0 keV. An extended source, very likely a galaxy cluster,
is also visible in the EPIC data. This cluster ($\alpha \sim$ 01:57:19
and $\delta \sim$ +37:56:43) is not visible in the previous {\sc
  Rosat} observations in this direction and it is at the edge of the
\chandra\ FOV. The source shows a slightly elongated circular
shape (with radius $R\sim 1'$) and a radially-decreasing profile of
surface brightness.

The spectrum of this extended source is well described  ($\chi^2 =
44.3$ with 33 d.o.f.) by an isothermal
optically-thin thermal plasma at $2.1_{-0.4}^{+0.5} \times 10^7$ K. The
mean metallicity is $Z=0.1^{+0.2}_{-0.1}$. The red-shift is consistent
with 0, but its upper limit  is $z=
0.1$. The unabsorbed X-ray flux per unit area in the $0.3-10$ keV
energy band is $\sim2.5 \times 10^{-14}$ erg cm$^{-2}$
arcmin$^{-2}$. These results suggest that the extended source is a
previously undetected cluster of galaxies. We note that the mean
metallicity of this new cluster is relatively low with respect to the
clusters with similar redshift presented in \citet{smk08}.

The final list of 145 point sources detected in the \xmm\ observation
is presented in Table\,7\footnote{on line.}. We
cross-matched the source list from \xmm\ with the one obtained from
the \chandra\ data, using a matching radius of 5 arcsec, and found 66
matches. For these sources, the Id of their \chandra\ counterpart is
given in Table\,7. In the \xmm\ source list
there are 23 sources\footnote{The \xmm\ sources with a counterpart in
  DLM94 within 5 arcsec are really 24, however, one of these, source
  69, has also a \chandra\ counterpart, which is not matched to the
  DLM94 counterpart (this being more than 7 arcsec away). Since the
  \chandra\ PSF is sharper, we consider the \chandra\ X-ray source
  position more accurate and thus unlikely associated with the star in
  DLM94, from now on, hence, source 69 will not be considered among
  the probable members of NGC~752.} with a counterpart from the
catalogue of NGC~752 by DLM94 (matching radius 5 arcsec): 18 are
likely cluster members, one is a possible member, and four are
unlikely cluster members (Table\,\ref{tab:xmm_vmatches}). Of the 19
likely or possible members, 12 were also detected in the
\chandra\ observations and the other seven are outside the
\chandra\ FOV.

\begin{table}[]
  \begin{center} 

\caption{\xmm\ sources with a visible counterpart from the study by
  DLM94.}

\leavevmode
\begin{tabular}{l|cccc|l}
\hline\hline
Id$_{\rm XMM}$ & Count rate & DLM94 Id & MC & $r$ & Id$_{\rm Chandra}$\\
~      & ks$^{-1}$ & ~ &  & arcsec & ~\\
\hline
15  & 0.81$\pm$0.13 & 654  & $+$ & 0.45 & 3\\
18  & 0.25$\pm$0.08 & 659  & $+$ & 2.36 & -\\
19  & 0.66$\pm$0.14 & 682  & $+$ & 2.32 & -\\
28  & 1.41$\pm$0.24 & 728  & $+$ & 2.24 & -\\
34  & 0.40$\pm$0.08 & 745  & $+$ & 1.95 & 32\\
40  & 0.47$\pm$0.09 & 772  & $+$ & 0.86 & 43\\
42  & 4.85$\pm$0.33 & 783  & $+$ & 1.26 & -\\
47  & 0.27$\pm$0.06 & 798  & $0$ & 2.77 & 70\\
57  & 0.60$\pm$0.11 & 824  & $+$ & 3.45 & 90\\
59  & 0.72$\pm$0.11 & 828  & $+$ & 1.22 & 93\\
64  & 8.82$\pm$0.30 & 849  & $+$ & 0.81 & 113\\
71  & 0.39$\pm$0.09 & 882  & $-$ & 4.44 & 140\\
75  & 1.48$\pm$0.23 & 890  & $+$ & 1.41 & -\\
80  & 0.54$\pm$0.10 & 901  & $+$ & 1.70 & 160\\
86  & 4.08$\pm$0.22 & 911  & $-$ & 0.34 & 176\\
94  & 0.31$\pm$0.08 & 935  & $+$ & 0.50 & 197\\
96  & 0.48$\pm$0.10 & 941  & $+$ & 1.76 & 201\\
98  & 7.50$\pm$0.30 & 950  & $+$ & 0.51 & 204\\
99  & 0.13$\pm$0.05 & 953  & $+$ & 0.69 & 207\\
101 & 1.92$\pm$0.20 & 958  & $-$ & 2.48 & 211\\
111\dag & 0.80$\pm$0.14 & 988  & $+$ & 4.40 & -\\
124 & 0.35$\pm$0.10 & 1023 & $+$ & 0.25 & -\\
128 & 0.87$\pm$0.16 & 1033 & $-$ & 2.11 & 258\\
\hline

\end{tabular}
\label{tab:xmm_vmatches}
  \end{center}

\dag X-ray source could be spurious

\end{table}

We searched for 2MASS counterparts to the \xmm\ sources using a
search radius of 5 arcsec and found a counterpart for  38
sources. As for the \chandra\ data, all sources with a visible
counterpart from DLM94 have also a 2MASS counterpart, so this leaves
15 \xmm\ sources with a 2MASS counterpart and no counterpart in
DLM94; of these, 3 were also detected by \chandra; of the other 
  12, 10 are outside the \chandra\ FOV, while two are within it
  (\xmm\ sources 58 and 65). Source 65 was caught by \xmm\ during the
  decay phase of a flare, which explains why it is not detected in the
  \chandra\ data. For source 58 there is no immediate explanation for
  this, since the light curve does not show evidence of a flare.
 No additional near-IR counterpart
  to the \xmm\ sources was found within the Point Source Reject Table
  of the 2MASS Extended Mission.

 As for \chandra\, we also searched for visible counterparts to the \xmm\ sources
  in the {\sc Nomad} catalogue (\citealp{zml+04}) and found 64 matches
  (within a search radius of 5 arcsec). Of these 64, 38 also have a
  2MASS counterpart. Of the 26 sources with only {\sc Nomad}
  counterpart, 24 lack $V$ magnitudes and have $B$ magnitudes $\ga 19$,
  suggesting that the great majority of them are background AGN. For
  the two sources with $V$ magnitude values, we checked their position
  in a $V$, $B-V$ diagram with respect to the isochrones by
  \cite{gbb+02} (see Sect.\,\ref{sec:res} for
  more details about the isochrones) and found that their colour  was
  incompatible with membership. 

For the point sources for which spectral and timing analysis was
carried out, the spatial regions for the sources and related local
background were defined as circles and annuli, respectively, with the
SAS task {\sc region} (to get the maximum signal to noise
ratio), and allowing shrinking of source regions in the case of nearby
sources. The related
response matrices and effective areas were then calculated.  For light
curves, we used an IDL procedure to bin photons in a grid of time
intervals with exposure times corrected for the Good Time Interval
selection. Light curves, images, and spectra, were created by
selecting events with {\sc pattern} $\le4$ for the PN camera, {\sc
  pattern} $\le12$ for the MOS cameras, and {\sc flag}$=0$ for
both. The data were screened by applying a count-rate limit on the
light curves (binned at 100 s) at high energies ($10-12$ keV for MOS
and $12-14$ keV for the PN camera), to eliminate the contamination by
soft proton flares. The count-rate limit is 0.35 counts per second for
the PN camera and 0.18 counts per second for the MOS cameras.

\section{Results}
\label{sec:res}

\subsection{Colour-magnitude diagrams}

\begin{figure*}[]
  \begin{center} \leavevmode 

        \epsfig{file=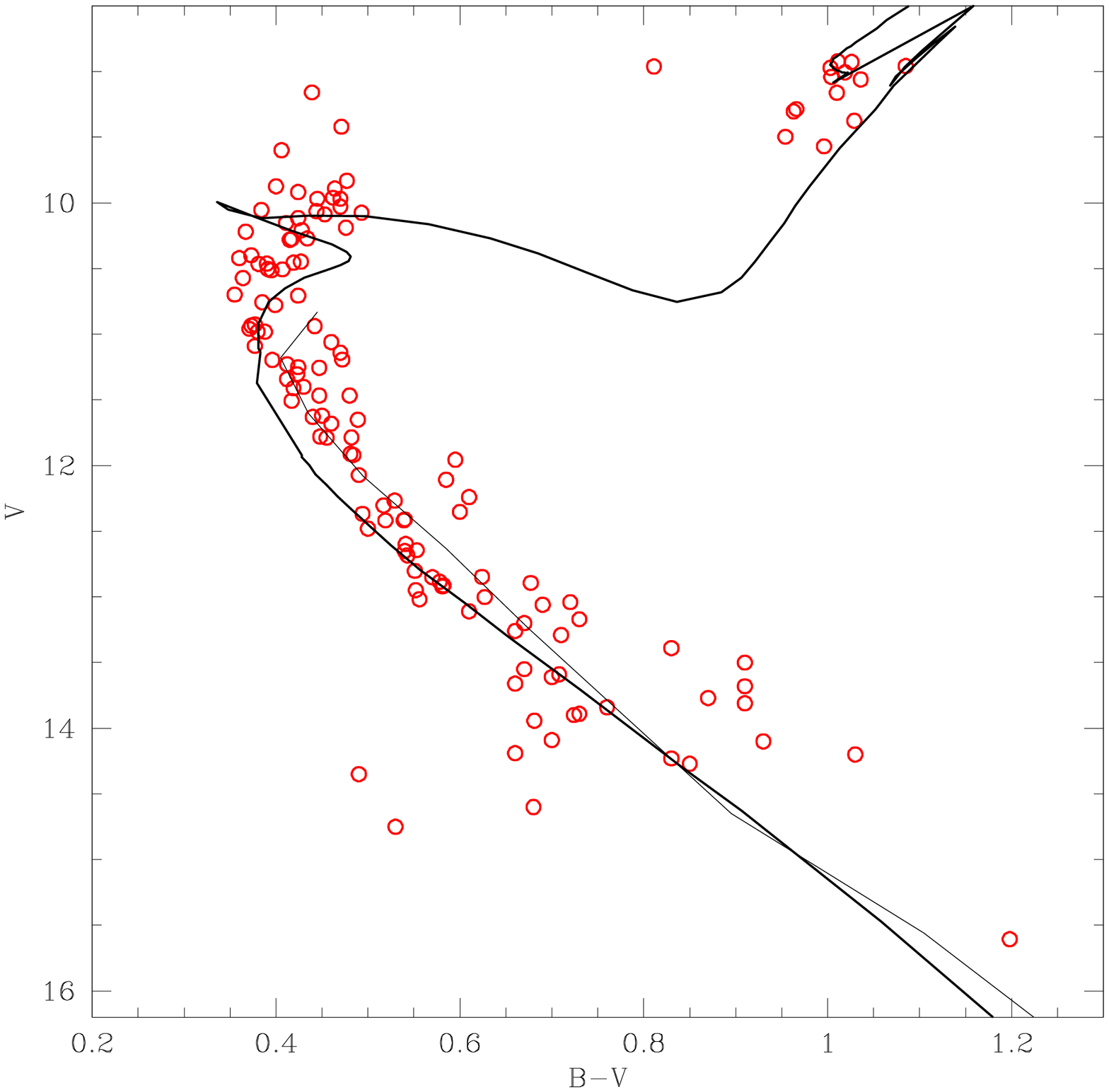, width=8.0cm, clip=}
	\epsfig{file=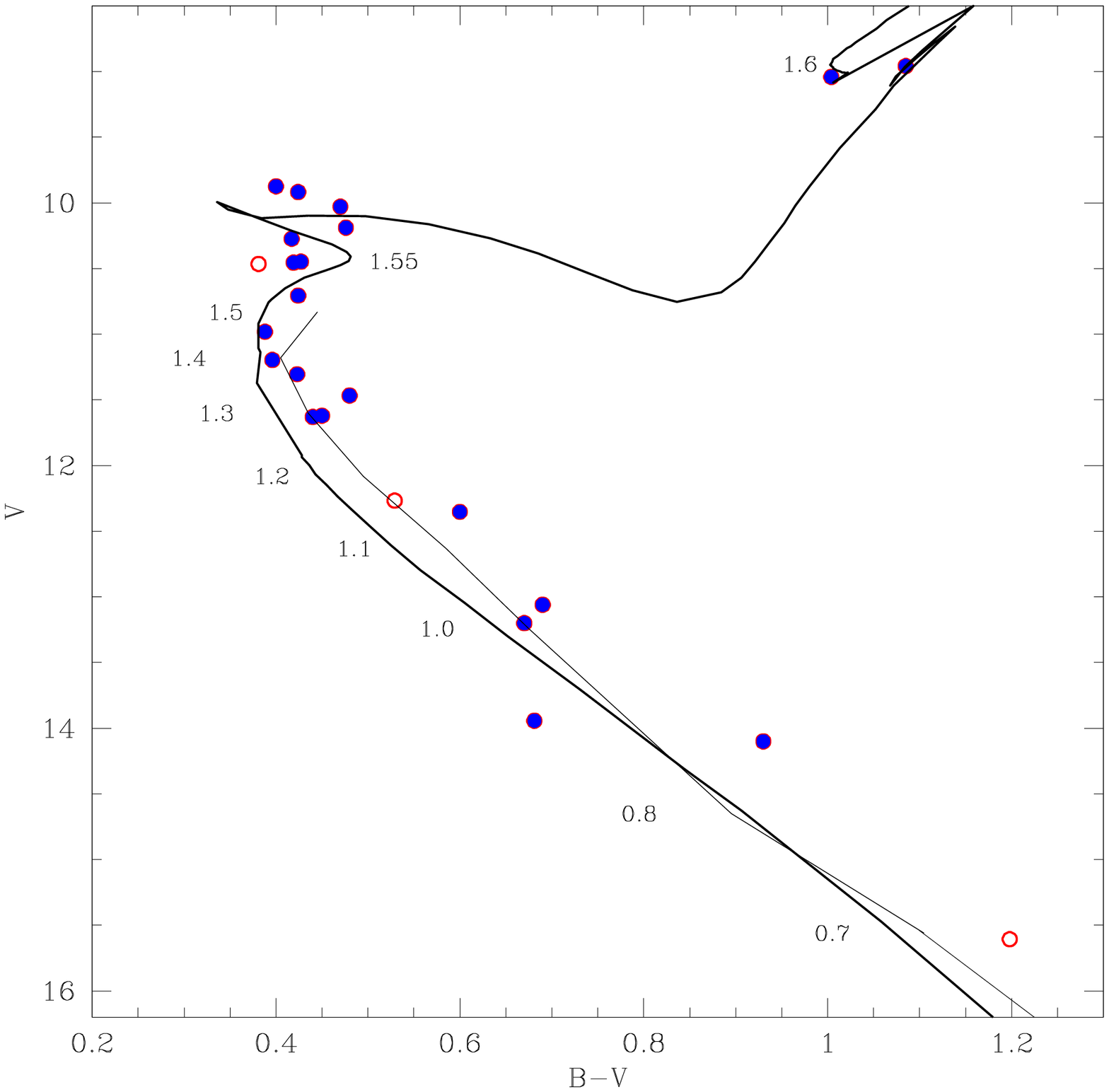, width=8.0cm, clip=}

\caption{Colour-magnitude diagrams for all known members of NGC~752
  ({\it Left} -- colours are from DLM94) and for all known members of
  NGC~752 within \chandra\ FOV ({\it Right}). In the right panel,
  filled circles represent sources with a
  \chandra\ counterpart. Continuous lines give the isochrone from
  Girardi et al. (2002) (thick line) and from Siess et al. (2000)
  (thin line). The theoretical isochrones are for 1.9 Gyr of age and
  $Z=0.01$ (shifted to the distance and reddening value of
  NGC~752). In the right panel the stars' masses along the theoretical
  isochrone are also indicated.}

    \label{fig:colmag}
  \end{center}
\end{figure*}

In the left panel of Fig.\,\ref{fig:colmag}, the colour-magnitude
diagram (CMD) for all probable and possible members of NGC~752 derived
using the data from DLM94 is shown (157 stars).  The right panel,
shows the cluster probable and possible members within the
\chandra\ FOV (24 stars), with the filled points indicating the stars
with an X-ray counterpart.  Since 21 \chandra\ sources have a
counterpart which is a probable (20) or possible member (1), this
implies that 88\% of (likely) cluster members within the FOV are
detected in the \chandra\ observation. In the \xmm\ FOV there are 47
objects classified as probable members of NGC~752 by DLM94 and, of
these, 19 are detected by \xmm\, which makes for a 40\% detection
rate.

In the figures the isochrones by \cite{gbb+02} and \cite{sdf00} are
also shown. These were computed for an age of 1.9 Gyr and assuming a
metallicity of $Z=0.01$. In the diagrams, they were shifted to the
distance and reddening value of NGC~752 (as given in
Sect.\,\ref{sec:ngc752}). As can be seen from Fig.\,\ref{fig:colmag},
there are only three X-ray undetected known members of the cluster,
and these three stars have very different masses ($\sim$ 1.5, 1.1, and
0.7 $M_{\sun}$). In particular we detect 5 out of the 6 known cluster
members with a mass between 0.8 and 1.2 $M_{\sun}$, that is, within
the mass range of our interest.

\begin{figure*}[]
  \begin{center} \leavevmode 

        \epsfig{file=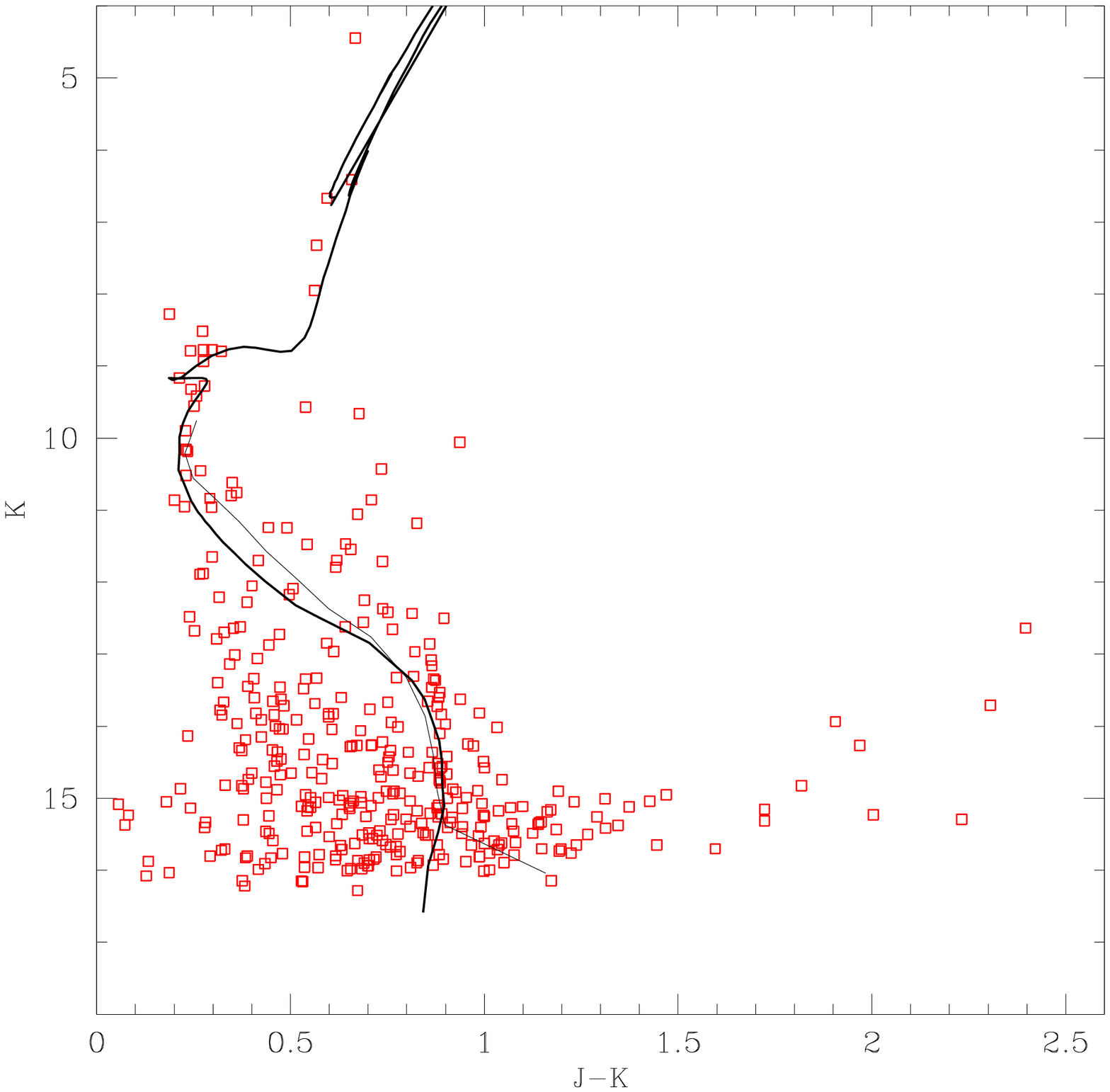, width=8.0cm}
	\epsfig{file=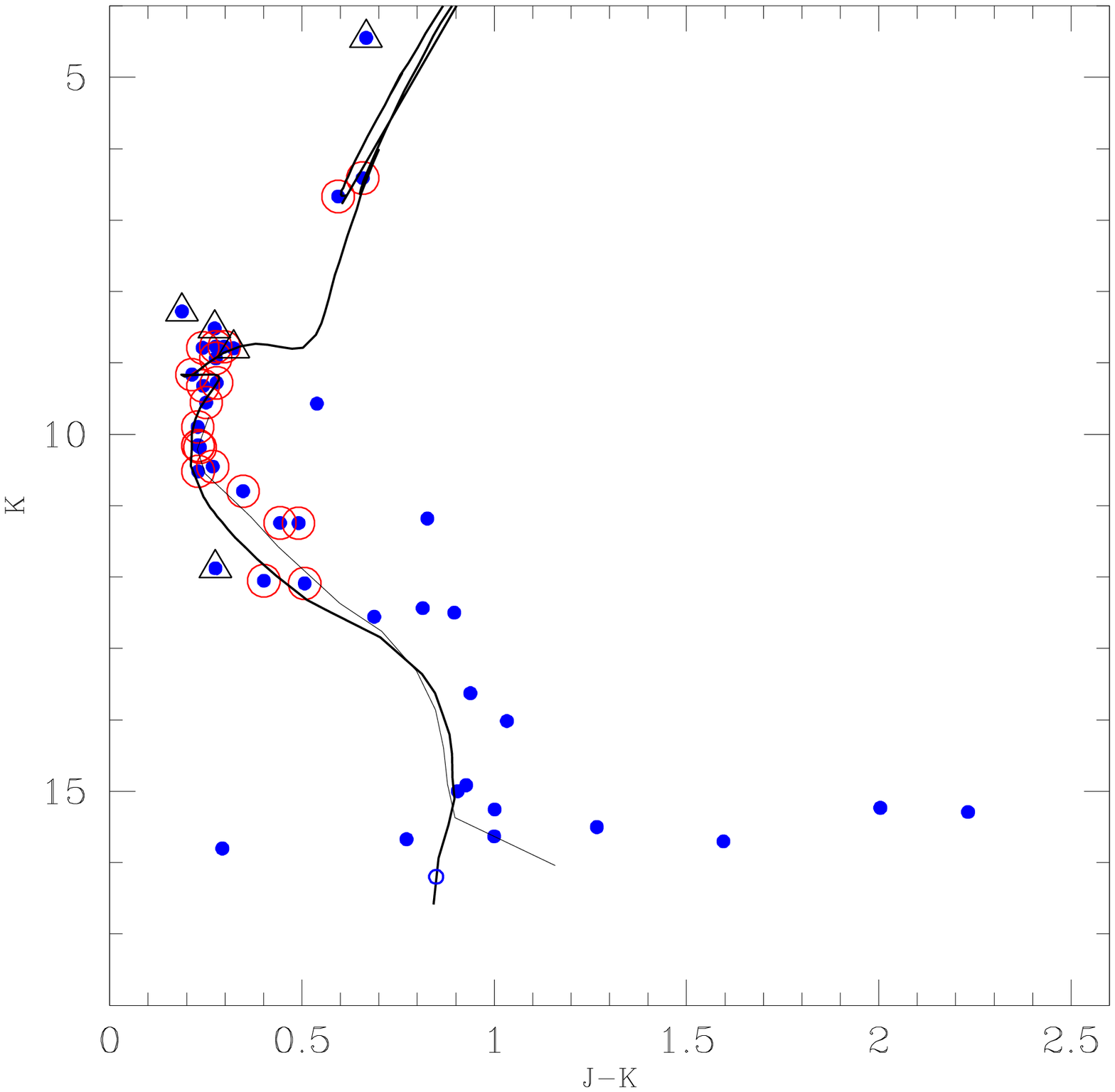, width=8.0cm}

\caption{Colour-magnitude diagrams for all 2MASS sources within
the \chandra\ FOV ({\it Left}) and for all \chandra\ X-ray sources  with a
2MASS counterpart ({\it Right}). Points with circles represent
known members of NGC~752,  while points with triangles are classified as
``likely non-member'' by DLM94. The empty point in the left digram
was matched by a source from 2MASS Extended Mission rather than from 2MASS PSC. 
Continuous lines are the isochrone from
Girardi et al. (2002) (thick line) and from Siess et al. (2000) (thin
line) as in the previous figure.}

    \label{fig:colmagIR}
  \end{center}
\end{figure*}

\begin{figure}
  \begin{center} \leavevmode 

        \epsfig{file=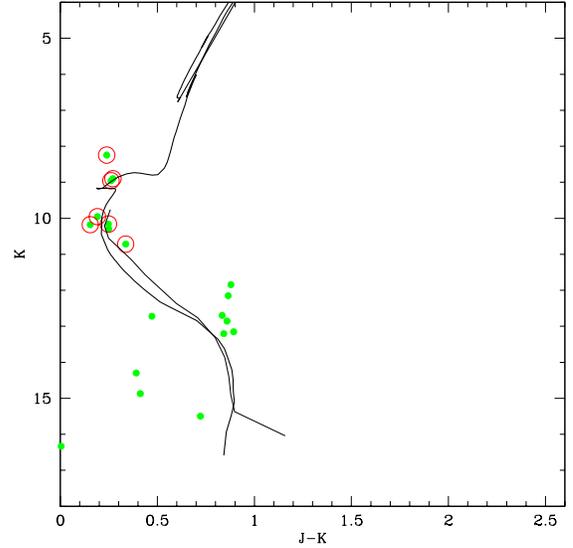, width=8.0cm}

\caption{Colour-magnitude diagram for the X-ray sources detected only
  by \xmm\ (i.e. not detected by \chandra), with a 2MASS
  counterpart. All the sources shown here a part from  two (\xmm\ ids 58
    and 65) fall
  outside the \chandra\ FOV. Symbols as in Fig.\,4.}

    \label{fig:colmagIRxmm}
  \end{center}
\end{figure}

The left panel of Fig.\,\ref{fig:colmagIR} shows the near-IR CMD
  for all 2MASS sources in the \chandra\ FOV,  while the
  right-panel shows the IR CMD for all the \chandra\ sources with a
2MASS counterpart, together with the isochrones from \cite{gbb+02} and
\cite{sdf00} (as for the CMDs in the visible).  The isochrones were
shifted to the distance and reddening value of NGC~752 using the
dereddening transformations $A_K/A_V = 0.108$ (\citealp{cox00}) and
$A_K = R_K E(J-K)$, with $R_K=0.66$ (\citealp{rl85}).  The IR CMD for
the  19 X-ray sources with a 2MASS counterpart detected only by
\xmm\ is shown in Fig.\,\ref{fig:colmagIRxmm}.

From Fig.\,\ref{fig:colmagIR}, one can see that 12 members of the
cluster, with masses approximately between 1.2 and 1.5 $M_{\sun}$, lie
very close to the isochrones at the turnoff. Sources 115 and 124 at
the top of the CMD are members of the cluster at the stage of core
helium burning and, with $M \sim 1.6 M_{\sun}$, they are the most
massive stars in our sample.  Beside the known cluster members, there
are three other stars very close to the isochrone turn-off, however,
DLM94 classify them as likely non-members because of their proper
motion.  For $K$ brighter than $K=12.5$ which corresponds to $M \ga
0.8M_{\sun}$, all the X-ray sources with a 2MASS counterpart have also
a counterpart in DLM94, apart from two sources to the right of the
isochrone: 182 and 185, which are at an angular separation of $\sim 8$
arcsec from each other. Source 185, a bright X-ray source, can be
identified using {\sc Simbad} with the bright star 915 in the study of
NGC~752 of \cite{platais91}, which, according to this study, has a
$0$\% probability of being part of the cluster (this star is omitted
from DLM94). The proper motion values given in the {\sc Nomad}
catalogue for this star confirm this: with $\mu_{\rm RA}=17.9 \pm 1.0$
mas/yr and $\mu_{\rm Dec}=-15.3 \pm 1.2$ mas/yr, source 185 is
unlikely to be a member of NGC~752, whose typical proper motion is
$\mu_{\rm RA}= 7.5$ mas/yr and $\mu_{\rm Dec}= -11.5$ mas/yr. No
counterpart is found for the fainter source 182 in the studies of
NGC~752, however, its counterpart in the {\sc Nomad} catalogue has
$\mu_{\rm RA}=8.2 \pm 9.0$ mas/yr and $\mu_{\rm Dec}=-88.5 \pm 9.0$
mas/yr, implying that also this star is very unlikely to be a member
of the cluster.

Away from the Galactic plane, the 2MASS Point Source Catalogue (PSC)
is complete down to $K = 14.3$ mag, in the absence of confusion, which
corresponds to $M \sim 0.45~M_{\sun}$ for this cluster. Nevertheless,
for $K < 12.5$ that is $M \ge 0.8~M_{\sun}$, cross-matching with the
2MASS catalogue allowed only one extra X-ray source to be identified
with respect to the studies by DLM94 and \cite{platais91} (source
182), and, in particular, no potential new cluster member was
identified with respect to DLM94. The sensitivity of the
\chandra\ data does not appear to be the limiting factor in
identifying new potential members with $M > 0.8~M_{\sun}$ within
2MASS, since the X-ray detection rate of cluster members is around
90\% for this mass range. This indicates that the list by DLM94 is a
complete census of cluster members within the \chandra\ FOV, for $M >
0.8~M_{\sun}$, and confirms the cluster to be deficient in solar-mass
stars with respect to the number of more massive members.
 
In the CMD of Fig.\,\ref{fig:colmagIR} (right panel), one can also see
11 objects fainter than $K=12.5$, not too far displaced from the
isochrones and thus with colours compatible with being members of the
cluster (Ids 21, 24, 35, 44, 87, 107, 136, 151, 223, 245, and 246). To
establish whether these sources could be possible members of the
cluster we checked their proper motion in the {\sc Nomad}
catalogue. Only sources 21, 44, 107, 136, 223, and 246 have proper
motion values compatible (within 4$\sigma$) with membership of
NGC~752. There is no counterpart in {\sc Nomad} for source 87 (within
3 arcsec).

In the CMD of Fig.\,\ref{fig:colmagIRxmm}, there are 9 objects not too
far displaced from the isochrones (\xmm\ sources ids: 2, 5, 11, 35,
58, 65, 95, 121, and 127), but only 2, 35, 58 and 127 have proper
motion compatible with cluster membership.

\subsection{Spectral and timing analysis for a few selected sources}

Of the \chandra\ sources with a 2MASS counterpart 12 have enough
source counts across the entire \chandra\ bandwidth for their spectra
to be analysed (count rate $>$ 0.3 ${\rm ks}^{-1}$).  The results of
the spectral and timing analysis for these 12 \chandra\ sources are
presented in Table\,\ref{tab:psfit}, with the possible or likely
members of the cluster in the first half of the Table.  We inspected
the light curve of all these sources and performed a
Kolmogorov-Smirnov test, which measures the maximum deviation of the
integral photon arrival times from a constant source model. The
results, in terms of constancy probability, are also given in the
Table.

All the spectral fits were
performed using an absorbed one-temperature plasma model. Initially,
we performed the fits with $N({\rm H})$ as a free parameter. All
12 fits, however, resulted in an essentially unconstrained value
for the absorbing column densities, so the fits were repeated fixing
 $N({\rm H})$ to its optically determined value of $2 \times
10^{20}{\rm cm}^{-2}$.  The
inability to tightly constrain the value of $N({\rm H})$ with these
ACIS-I data is not unexpected, since the instrument's energy band
extends only to 0.2 keV at the low energies, and the values of $N({\rm
H})$ here are rather low.  

 As shown in Table\,\ref{tab:psfit}, all the four established cluster members have
temperature $kT$ around 0.5 $-$ 0.7 keV; indeed, all the sources
except 180 and 223 have a similar (relatively) low plasma temperature,
consistent with being stellar sources.  Source 223 has $kT = 1.9$ keV,
however, this source underwent a strong flare during the observation:
its count rate increased impulsively by more than a factor of 30 and
then decreased exponentially over 18 ks. A spectral fit to the source
data during the quiescent time yields $kT = 0.6 \pm 0.2$ keV, similar
to the value of the other sources.  The spectrum of source 180 is too
hard for a normal star, thus this source is very likely a background
AGN. Indeed its spectrum can be equally well fitted with an absorbed
power law with a spectral index $\gamma = 1.9 \pm 0.1$
(null-hypothesis probability $P=0.96$). Its 2MASS colours are
consistent with this source being in the background.

Source 185 underwent a flare at about 30 ks from the start of
the observations; during the flare, the source's count rate increased
by a factor $\sim 5$. The spectrum of this source integrated over the
entire exposure is not well modeled by an absorbed one-temperature
plasma model. We subdivided the spectral data for this source in two
time intervals, one during the flare and one during the star's
quiescent activity. During the quiescent phase the spectrum is well
fitted ($P = 0.14$) by an absorbed two-temperature model with $kT_1 =
0.35 \pm 0.04$ keV, $kT_2=0.97\pm 0.05$ keV, $EM_1=0.7 \times
10^{52}$~cm$^{-3}$, $EM_2=0.53\times 10^{52}$~cm$^{-3}$, and $Z=0.22
\pm 0.05 Z_{\sun}$; the value of $N({\rm H})$ is also in this case not
well constrained: $(0.0 \pm 0.7) \times 10^{20}$ \cmtwo. During the
flare, the value of $N({\rm H})$, $kT_1$, $EM_1$, and $Z$ remain
essentially unchanged, but $kT_2$ increases to $1.9 \pm 0.2$ and
$EM_2$ to $1.3 \times 10^{52}$~cm$^{-3}$ (null-hypothesis probability
of the fit $P = 0.35$).

In Table\,\ref{tab:xmm_psfit} the results of the spectral analysis for
\xmm\ sources with a 2MASS counterpart and a count rate greater than 1.5
cts/ks are presented (these are also 12 sources). For each source, the
fit was performed jointly to PN, MOS1, and MOS2 spectral data.  As
in the table, for the stellar sources for which both \chandra\ and
\xmm\ data are available, the best fit values of the plasma temperature
and the emission measure are in good agreement between the two data
sets.

The typical value of the plasma temperature for the possible or likely
members of NGC~752 is 0.7~keV. This value is similar to the
plasma temperatures derived for members of the (0.8 Gyr old) Hyades
cluster and somewhat lower than the typical plasma temperature for the
much younger Pleiades ($\sim$ 0.08 Gyr). \cite{ssp+94} report the
results of 2T temperature fits to the {\sc Rosat} spectral data of 11
members of the Hyades. The mean value of the highest temperature is
1.0 keV (all values) or 0.8 keV if one excludes one outlier with
$kT_2=3.4$ keV. If one computes the emission-measure-weighted mean
temperature for each star,  the mean value for the 11 Hyades
members is 0.6 keV.  For the Pleiades, \cite{gcs95} performed 2T
temperature fits to the spectral data of 16 members and provides a
value for the mean plasma temperature for each star. The mean of these
16 values is 0.9 keV.

\begin{table*}
  \begin{center}

\caption{Best-fit spectral parameters for \chandra\ sources with a 2MASS
counterpart and a count rate greater than 0.3 cts/ks$^{(a)}$.}
\leavevmode
     \begin{tabular}{l|cccccccc|l}
\hline\hline

Src Id &   $kT$ & $Z$ & $E\!M$ & $\chi^2/{\rm d.o.f.}$ & $P$ & $F_{\rm
     X}$ & $L_{\rm X}$ & $C_{\rm KS}^{(b)}$ & Note \\
\hline
~ & keV &  $Z_{\odot}$ & $E\!M_{52}^{(c)}$ & ~ & ~ & $F_{-13}^{(d)}$  &
     $L_{29}^{(e)}$ & \% & \\
\hline

3   &   0.49 $\pm$ 0.16 & 0.30 (froz.) & 0.56 $\pm$ 0.50 & 0.99 & 0.45 & 0.02 & 0.60 & 38 & Likely memb. (DLM94)\\
43  &   0.76 $\pm$ 0.07 & 0.30 (froz.) & 0.64 $\pm$ 0.20 & 0.58 & 0.86 & 0.03 & 0.81 & 0.0 & Likely memb. (DLM94)\\
113$\clubsuit$  &   0.71 $\pm$ 0.02 & 0.24 $\pm$ 0.05 & 10.72 $\pm$ 3.26 & 1.43 & 0.01 & 0.48 & 11.65 & 0.6 & Likely memb. -- SB2
(DLM94)\\
204 &   0.64 $\pm$ 0.02 & 0.61 $\pm$ 0.27 & 3.61 $\pm$ 2.88 & 1.30 & 0.07 & 0.30 & 7.37 & 3.0 & Likely memb. -- SB2 (DLM94)\\
223\dag &   1.94 $\pm$ 0.27 & 0.30 (froz.) & 2.08 $\pm$ 0.36 & 0.95 & 0.54 & 0.11 & 2.53 & 0.0 & Possible memb. (IR colours)\\
\hline
151 &   0.75 $\pm$ 0.06 & 0.07 $\pm$ 0.03 & 5.28 $\pm$ 2.45 &
0.82 & 0.76 & 0.14 & - & 0.2 & Unlikely memb. ({\sc Nomad} prop mot.)\\
176 &   0.63 $\pm$ 0.03 & 0.71 $\pm$ 0.48 & 1.62 $\pm$ 1.98 & 1.17 & 0.21 & 0.16 & - & 4.0 & Unlikely Memb. (DLM94)\\
180 &   3.73 $\pm$ 0.38 & 0.30 (froz.) & 32.26 $\pm$ 2.27 & 0.98 & 0.56 & 2.15 & - & 13 & Likely background AGN\\
182 &   0.71 $\pm$ 0.04 & 0.09 $\pm$ 0.03 & 9.87 $\pm$ 3.55 & 1.18 & 0.16 & 0.27 & - & 0.0 & Unlikely memb. ({\sc Nomad} prop. mot.)\\
185\dag &   0.73 $\pm$ 0.02 & 0.06 $\pm$ 0.01 & 107.27 $\pm$
10.71 & 1.72 & 0.00 & 3.03 & - & 0.0 & Not a member$\diamondsuit$\\
211 &   0.51 $\pm$ 0.09 & 0.30 (froz.) & 1.33 $\pm$ 0.60 & 1.13 & 0.32 & 0.06 & - & 0.07 & Unlikely Memb. (DLM94)\\
258 &   0.30 $\pm$ 0.05 & 0.30 (froz.) & 1.63 $\pm$ 1.07 & 0.49 & 0.90 & 0.05 & - & 0.01 & Unlikely Memb. (DLM94)\\
\hline

\end{tabular}
    \label{tab:psfit}
  \end{center}

$^{(a)}$ the table is
divided in two parts, with likely or possible members of the cluster
in the first half

$^{(b)}$  probability of constancy from the Kolmogorov-Smirnov test for the
sources' light curves

$^{(c)}$ $E\!M_{52} = 10^{52}$~cm$^{-3}$

$^{(d)}$ $F_{-13}= 10^{-13}$~\ecms

$^{(e)}$ $L_{29} = 10^{29}$~\es (assuming $d=430$ pc)

$\clubsuit$ source affected by $\sim 1$\% pileup

\dag flare

$\diamondsuit$ This source corresponds to source 915 in the
study by Platais\,(1991), which is classified as non-member.

\end{table*}

\begin{table*}
  \begin{center}

\caption{Best-fit spectral parameters for \xmm\ sources with a 2MASS
  counterpart and a count rate greater than 1.5 cts/ks (symbols and
  units as per Table 3.)}

\leavevmode
     \begin{tabular}{ll|cccccccc|l}
\hline\hline
Id$_{\rm xmm}$ & Id$_{\rm chandra}$ & $kT$ & $Z$ & $E\!M$ & $\chi^2/{\rm d.o.f}$ & $P$ & $F_{\rm X}$ & $L_{\rm X}$ & $C_{\rm KS}$ & Note \\
\hline
~ & ~ & keV &  $Z_{\odot}$ & $E\!M_{52}$ & ~ & ~ & $F_{-13}$  &
     $L_{29}$ & \% & \\
\hline

42   & -   & 0.94 $\pm$ 0.09 & 0.07 $\pm$ 0.02 & 11.55 $\pm$ 2.57 & 1.16 & 0.17 & 0.34 & 8.30 & 0.0  &  Likely memb. (DLM94)\\
64   & 113  & 0.67 $\pm$ 0.01 & 0.45 $\pm$ 0.08 & 6.13 $\pm$ 1.93 & 1.12 & 0.16 & 0.42 & 10.06 & 0.0  &  Likely memb. (DLM94)\\
98   & 204 & 0.65 $\pm$ 0.02 & 0.68 $\pm$ 0.21 & 3.84 $\pm$ 2.10 & 1.13 & 0.21 & 0.36 & 8.63 &  44 &  Likely memb. (DLM94)\\
127 &  -  & 0.71 $\pm$ 0.06 & 0.30 (froz.) & 1.47 $\pm$ 0.39 & 0.78 & 0.70 & 0.07 & 1.69 &  0.5 &  Possible memb. (IR colours)\\
\hline
5    & -   & 0.76 $\pm$ 0.06 & 0.09 $\pm$ 0.03 & 8.15 $\pm$ 2.57
& 0.91 & 0.69 & 0.23 & - & 3.0  & Unlikely memb. ({\sc Nomad}
prop. mot.)\\
11    & - & 0.63 $\pm$ 0.03 & 0.30 (froz.) & 4.77 $\pm$ 0.50 & 1.48 & 0.03 & 0.24 & - &  7.0  & Likely background
AGN\ddag\\
76  & 151 & 0.71 $\pm$ 0.05 & 0.30 (froz.) & 1.68 $\pm$ 0.30 & 1.14 & 0.21 & 0.09 & - & 0.1  &  Unlikely memb. ({\sc Nomad}
prop. mot.)\\
86   & 176 & 0.57 $\pm$ 0.03 & 0.30 (froz.) & 4.94 $\pm$ 0.41 & 1.76 & 0.01 & 0.23 & - &  0.0  & Unlikely Memb. (DLM94)\\
87   & 180 & 2.24 $\pm$ 0.09 & 0.30 (froz.) & 39.50 $\pm$ 1.12 & 2.52 & 0.00 & 2.14 & - & 0.0   & Likely background AGN\ddag\\
89  & 185 & 0.67 $\pm$ 0.01 & 0.11 $\pm$ 0.01 & 66.71 $\pm$ 4.46 & 1.20 & 0.02 & 1.91 & - & 0.0   & Not a member\\  
101   & 211 & 0.66 $\pm$ 0.07 & 0.30 (froz.) & 1.94 $\pm$ 0.35 & 1.18 & 0.20 & 0.10 & - &  44  & Unlikely Memb. (DLM94)\\
121  & -   & 0.99 $\pm$ 0.10 & 0.30 (froz.) & 2.86 $\pm$ 0.87 & 1.72 & 0.00 & 0.14 & - &  9.0  &  Likely background
AGN\ddag\\

\hline

\end{tabular}
    \label{tab:xmm_psfit}
  \end{center}
\ddag A Better fit to this source's spectrum is obtained with an absorbed
    power law model
\end{table*}

\subsection{X-ray luminosity}

For all the members of NGC~752 which have an X-ray counterpart (21
sources detected by \chandra\ plus 7, outside \chandra\ FOV, detected
by \xmm) we provide an estimate of the source X-ray luminosity in
Table\,\ref{tab:lx}. The estimates for a star's mass and bolometric
luminosity were obtained using the isochrones by \cite{gbb+02} (age
$=$ 1.9 Gyr, Z = 0.01). The count rate to flux conversion factor was
obtained using the {\sc pimms} software at {\sc Heasarc}, and assuming
for the emitting plasma a Raymond-Smith model with $kT = 0.7$ keV and
$Z=0.2~Z_{\sun}$; the absorbing column density was set to $N({\rm H})
= 2 \times 10^{20}$ \cmtwo, as derived from the cluster $E(B-V)$. The
conversion factor is $7.6 \times 10^{-15}{\rm erg~cm^{-2}
  s^{-1}~ks^{-1}}$ for \chandra\ ACIS-I and $5.1 \times 10^{-15}{\rm
  erg~cm^{-2} s^{-1} ks^{-1}}$ for the \xmm\ EPIC PN and MOS detectors
combined\footnote{The conversion factor for the count rate of the
  \xmm\ sources is the weighted mean of the conversion factors given
  by {\sc pimms} for MOS1, MOS2 and PN, with weights 1,1, and 3
  respectively. This is because in the source detection algorithm for
  the \xmm\ merged file events the efficiency of PN was taken to be 3
  times that of the MOS detectors.}.

Using the above conversion factors one can compare the fluxes from
\chandra\ and \xmm\ sources. The correlation coefficient for the  66
commonly detected sources is 1.06 $\pm$ 0.04, if one restricts the
analysis to the 12 members of NGC~752 commonly detected (for which our
choice of {\sc pimms} input parameters is appropriate), the
correlation coefficient is 0.99 $\pm$ 0.06. As can be seen from
Table\,\ref{tab:lx}, there is good agreement between the \chandra\ and
\xmm\ luminosity values for these 12 sources.  Note also that for the
sources in Table\,\ref{tab:lx} for which a spectral analysis was
possible, the X-ray luminosity given here compares well with the
values derived from the spectral fits.

Among the cluster members, most of the sources (21) are concentrated
at the main sequence turn-off in the CMD and have masses around 1.3
and 1.5 solar masses. Most of these sources have a similar X-ray
luminosity, between 0.1 and $0.7 \times 10^{29}$ \es; the exceptions,
with $L_{\rm X}\ga 10^{29}$ \es, are the two binary systems (DLM94),
113 and 204, the fast rotators (DLM94) 28$_{\rm xmm}$ and 111$_{\rm
  xmm}$ (this last one is also a binary system), and 75$_{\rm
  xmm}$. The two cluster members at the stage of core helium burning
(115 and 124) have the same X-ray luminosity. There are only six
sources within the mass range of our interest ($0.8-1.2~M_{\sun}$)
among the likely cluster members; of these, five sources have
luminosity in the range 0.1-0.4 $\times 10^{29}$ \es; the other one,
source 42$_{\rm xmm}$, is significantly brighter. This star does not
show evidence of being a close binary nor a fast rotator;
\cite{bv2001} classified it as F5$-$F7 and report that no Ca\,{\sc ii}
H\&K are visible in the low-resolution spectra.  For source 93, the
X-ray luminosity derived from the \chandra\ data is $\sim 6$ times
lower than that derived from the \xmm\ data, however, this source's
light curve shows evidence that the source was caught by \xmm\ during
the decay phase of a flare.

In Table\,\ref{tab:lx} we also provide an estimate of the X-ray
luminosity of 11 sources which have no counterpart among the cluster
members listed in DLM94, but which have 2MASS colours and proper
  motion (in {\sc Nomad}) compatible
  with cluster membership; in addition their  X-ray
spectrum or hardness ratio (HR) is consistent with the source being a
star. 
These 11 cluster member candidates
have all a lower mass than those from DLM94, with their position in the
IR CMD consistent with masses $< 0.8 M_{\sun}$.

\begin{figure}[]
  \begin{center} \leavevmode 

        \epsfig{file=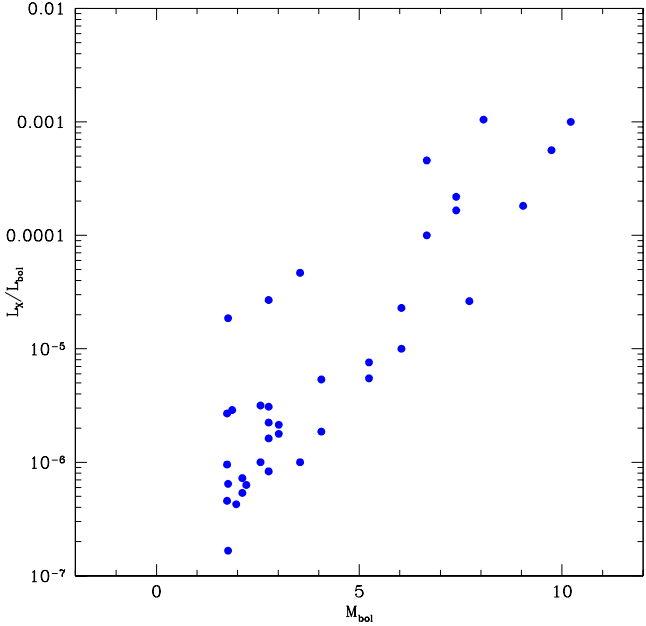, width=9.0cm, clip=}

\caption{$L_{\rm X}/L_{\rm bol}$ vs. bolometric magnitude for the
  sources in Table 5. Note the deficiency of faint members; a couple
  of M stars at the ``saturation'' level is also noticeable.}

    \label{fig:LxLbol}
  \end{center}
\end{figure}

In Fig.\,\ref{fig:LxLbol} the value of $L_{\rm X}/L_{\rm bol}$ from
the table are plotted against a star's bolometric magnitude, $M_{\rm
  bol}$. This plot can be directly compared with the plot of $L_{\rm
  X}/L_{\rm bol}$ vs $M_{\rm bol}$ for the Hyades in figure 15 of
\cite{ssk95}.  The comparison shows that for stars with bolometric
magnitude $M_{\rm bol} < 8$ (corresponding to $M\ga 0.5 M_{\sun}$) the
sensitivity of the \chandra\ observation in terms $L_{\rm X}/L_{\rm
  bol}$ is such that we do not expect to lose a significant fraction
of the cluster population within the \chandra\ FOV (possibly a couple
of members), consistent with a \chandra\ detection rate of 88\%. From
the comparison of our Fig.\,\ref{fig:LxLbol} and Fig. 15 in
\cite{ssk95}, the deficiency of low mass stars with respect to cluster
members at the main-sequence turn-off in NGC~752 is again
apparent. Comparison of the two figures shows also that this is not an
effect of the sensitivity of the \chandra\ observation. This result
and the fact that cross-matching with the 2MASS catalogue did not
identify any new candidate members with $M > 0.8~M_{\sun}$ beside
those already listed in DLM94, confirms that (within the \chandra\ FOV
where our sensitivity is higher) the sample by DLM94 is complete for
$M > 0.8~M_{\sun}$ and NGC~752 is intrinsically deficient in low mass
stars.
 
\begin{figure}[]
  \begin{center} \leavevmode 

        \epsfig{file=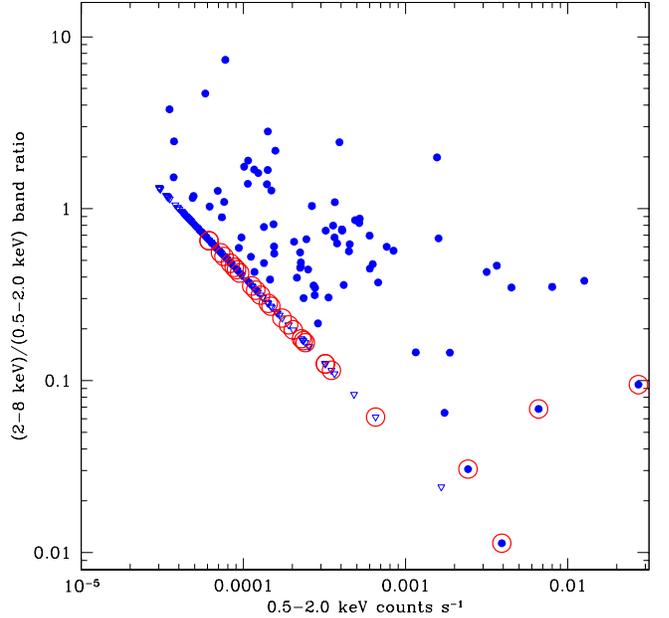, width=9.0cm}

\caption{Hardness ratio of all \chandra\ sources. Points with
circles represent known stars in the field. 
Empty triangles give indicative upper limits for
the sources which were not detected in the hard band.}

\label{fig:HR} \end{center}
\end{figure}

In Fig.\,\ref{fig:HR}, we show the scatter-plot of all the
\chandra\ sources as a function of their HR. The HR values were
obtained by cross-matching the 262 sources detected in the $0.5-2$ keV
band with the sources detected in the $2-8$ keV band, with similar
detection parameters (139 sources); upper limits are given for the
sources which were not detected in the hard band. All the catalogued
stars in the field have HR below $\sim$ 0.2 or an upper limit because
they were not detected in the hard band at all.

\begin{table*}[]
  \begin{center} 

\caption{X-ray luminosity of sources with a 2MASS counterpart and
  which are likely or possible members of NGC~752$^{(a)}$.}

\footnotesize
\leavevmode
\begin{tabular}{llccccccl}
\hline

Id$_{\rm chandra}$ & Id$_{\rm xmm}$ &  Rate & Mass  & $L_{\rm
  X}$(${\rm chandra}$) & $L_{\rm X}$(${\rm xmm}$) & $\log(L_{\rm
  bol}/L_{\sun})$ & $\log(L_{\rm X}/L_{\rm bol}) $ &  Notes$^{(b)}$\\
\hline
   &  & $ks^{-1}$ & $M_{\sun}$  &  $10^{29}$ \es &  $10^{29}$ \es & &
& \\
\hline
3     &  15 & 0.32   &    1.38   &  0.54    & 1.38 & 0.79    &     -5.65 & \\
17    &  -  & 0.08   &    1.11   &  0.13    &   -  & 0.27    &     -5.73 & \\
32    &  34 & 0.23   &    1.57   &  0.39    & 0.45 & 1.19    &     -6.19 & SB1 \\
43    &  40 & 0.35   &    1.57   &  0.59    & 0.53 & 1.20    &     -6.02 & ~\\
70    &  47 & 0.19   &    1.55   &  0.32    & 0.30 & 1.05    &     -6.14 & SB2 \\
71    &  -  & 0.12   &    1.38   &  0.20    &   -  & 0.79    &     -6.08 & ~\\
86    &  -  & 0.13   &    1.57   &  0.22    &   -  & 1.11    &     -6.37 & ~ \\
90    &  57 & 0.2    &    1.32   &  0.34    & 0.68 & 0.69    &     -5.75 & ~ \\
93    &59\dag& 0.08  &    0.91   &  0.13    & 0.81 & -0.20   &     -5.26 & SB1\\
113   &  64 & 6.6    &    1.57   &  11.10   & 9.95 & 1.19    &     -4.73 & SB2\\
115   &  -  & 0.09   &    1.59   &  0.15    &   -  & 1.74    &     -7.15 & SB1\\
116   &  -  & 0.06   &    1.57   &  0.10    &   -  & 1.19    &     -6.78 & SB1, ROT\\
117   &  -  & 0.23   &    1.11   &  0.39    &   -  & 0.27    &     -5.27 & ~\\
124   &  -  & 0.09   &    1.59   &  0.15    &   -  & 1.74    &     -7.15 & ~\\
145   &  -  & 0.14   &    1.55   &  0.24    &   -  & 1.05    &     -6.27 & ~\\
160   &  80 & 0.17   &    1.43   &  0.29    & 0.61 & 0.87    &     -6.00 & ROT\\
184   &  -  & 0.11   &    0.91   &  0.18    &   -  & -0.20   &     -5.12 & ~\\
197   &  94 & 0.24   &    1.32   &  0.40    & 0.35 & 0.69    &     -5.67 & ~\\
201   &  96 & 0.15   &    1.52   &  0.25    & 0.54 & 1.01    &     -6.20 & ROT\\
204   &  98 & 3.92   &    1.38   &  6.59    & 8.46 & 0.79    &     -4.57 & SB2\\
207   &  99 & 0.07   &    1.20   &  0.12    & 0.15 & 0.48    &     -6.00 & SB1\\
$\sharp$& 18& 0.25   &	  1.57	 &$\sharp$  & 0.28 & 1.20    &     -6.34 & ~\\
$\sharp$& 19& 0.66   &	  1.38	 &$\sharp$  & 0.74 & 0.79    &     -5.51 & ~\\
$\sharp$& 28& 1.41   &	  1.58	 &$\sharp$  & 1.59 & 1.15    &	   -5.54 & ROT\\
$\sharp$& 42& 4.85   &	  1.20	 &$\sharp$  & 5.47 & 0.48    &     -4.33 & ~\\
$\sharp$& 75& 1.48   &    1.57	 &$\sharp$  & 1.67 & 1.20    &     -5.57 & ~\\
$\sharp$&111& 0.80   &	  1.43	 &$\sharp$  & 0.90 & 0.87    &     -5.50 & SB1, ROT\\
$\sharp$&124& 0.35   &	  1.38	 &$\sharp$  & 0.39 & 0.79    &     -5.79 & SB1\\
\hline
21      & - & 0.08   &     0.35  &  0.13    &   -  & -1.72 &      -3.74 & \\
44      & - & 0.13   &     0.25  &  0.22    &   -  & -2.00 &      -3.25 & \\
87      & 56& 0.15   &     0.20  &  0.25    & 0.70 & -2.19 &      -3.00 & \\
107     & - & 0.04   &     0.55  &  0.07    &   -  & -1.19 &      -4.58 & \\
136     & - & 0.16   &     0.79  &  0.27    &   -  & -0.52 &      -4.64 & \\
223\dag & - & 1.15   &     0.50  &  1.93    &   -  & -1.33 &      -2.98 & \\
246     & - & 0.07   &     0.79  &  0.12    &   -  & -0.52 &      -5.00 & \\
$\sharp$&2  & 0.66   &	   0.60	 &$\sharp$  & 0.74 & -1.06 &	  -3.66 & \\
$\sharp$&35 & 0.5    &	   0.60	 &$\sharp$  & 0.56 & -1.06 &	  -3.78 & \\
  -     &58 & 0.59   &	   0.70	 &   -      & 0.67 & -0.77 &	  -4.00 & \\
$\sharp$&127& 2.7    &	   0.70	 &$\sharp$  & 3.05 & -0.77 &  	  -3.34 & \\
\hline
\end{tabular}
    \label{tab:lx}
  \end{center}

$^{(a)}$ The first 28
  sources are listed in DLM94 and classified as likely or possible
  members of NGC~752, the last 11 sources are not listed in DLM94 but
  have 2MASS colours and {\sc Nomad} proper motion consistent with the
  star being a member of NGC~752

$^{(b)}$ Notes are from DLM94

- source was not detected

$\sharp$ source was outside FOV

\dag flare

\end{table*}

\subsubsection{Comparison with the {{\sc Rosat}} observation}

As mentioned in Sect.\,\ref{sec:ngc752}, NGC~752 was observed in the
X-ray with {\sc Rosat} by \cite{bv96} who identified seven cluster
members; of these, 3 have \chandra\ and \xmm\ counterparts, one has
only \xmm\ counterpart (being outside the \chandra\ FOV) and the other
three are outside the \xmm\ FOV. Although source BV6 is identified
with a member of NGC~752 by \cite{bv96}, the corresponding source in
\chandra\ (Id 131) does not match any of the sources in DLM94 nor in
2MASS within a search radius of 3 arcsec. The \chandra\ X-ray source
is $\sim 8$ arcsec away from the member of NGC~752 with DLM94 Id 868
(and the corresponding 2MASS source), which explains why \cite{bv96}
found the associations with a cluster member for the {\sc Rosat}
source. Given the accuracy of \chandra\ and 2MASS positions we
discarded this X-ray source as possible counterpart of the cluster
member. Sources BV11 and BV16 correspond to the \chandra\ sources 204
and 113 respectively and their {\sc Rosat} luminosity is within 50\%
from the values given in Table\,\ref{tab:lx}. BV29 corresponds to
42$_{\rm xmm}$ and its {\sc Rosat} luminosity is about a factor of 3
higher than that derived here. BV30, BV45 and BV46 are outside
\xmm\ FOV, their colours corresponds to star masses of 1.0, 1.6, and
0.9 $M_{\sun}$ respectively.  The X-ray luminosity of the two
solar-mass stars BV30 and BV46 ($L_{\rm X} = 2.3 \times
10^{30}$\es\ and $L_{\rm X} = 1.3 \times 10^{30}$\es) are
significantly higher than the X-ray luminosity of any of the
solar-mass stars in our sample and correspond to $\log(L_{\rm
  X}/L_{\rm bol})$ values of $-3.4$ and $-3.3$ respectively, that is
$L_{\rm X}/L_{\rm bol}$ a factor of 10 or higher than the values in
our sample or in the Hyades (\citealp{ssk95}), for this mass range,
suggesting that these two sources were probably undergoing a flare
during the {\sc Rosat} observation.

\section{Discussion and conclusions}
\label{sec:disc}

\begin{figure}[]
  \begin{center} \leavevmode

        \epsfig{file=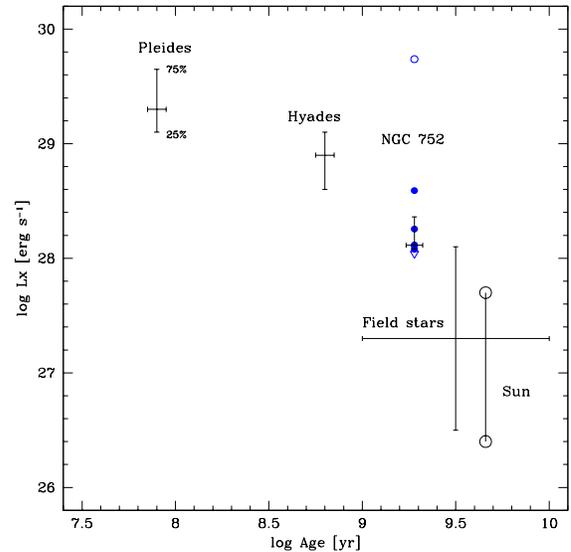, width=8.0cm}

\caption{X-ray luminosity of members of NGC~752 with mass between 0.8
  and 1.2 $M_{\sun}$ (filled points for \chandra\ detected stars
    and empty point for the one detected by \xmm\ outside
    \chandra\ FOV), compared with median luminosities of stars within
  the same mass range in the Pleiades, Hyades and a sample of field
  stars. The triangle indicates the upper limit for star Id 790 from
  DLM94. Error bars indicate the median value with the size of the
  vertical bar indicating the 25\% and 75\% quantiles of the
  distribution (see text). The line at age 4.5 Gyr connects the minimum
  and maximum of the Sun.}

\label{fig:lxvage} 
\end{center}
\end{figure}

The aim of this paper is to determine the level of stellar activity of
solar type stars at an intermediate age between that of the Hyades and
those of field stars.  In Fig.\,\ref{fig:lxvage}, we compare the
luminosities of members of NGC~752 with masses in the range
$0.8-1.2~M_{\sun}$ with the luminosity distributions of the Pleiades,
the Hyades and the field stars. These three points are from
\cite{micela02} and refer to stars within the same mass range
($0.8-1.2~M_{\sun}$). As summarised in Table\,\ref{tab:lx}, in NGC~752
we detected in the X-ray 6 cluster members within this mass range (the
points in the figure), 5 stars within the \chandra\ FOV and one
outside (significantly brighter than the others) detected by \xmm. For
stars detected both by \chandra\ and \xmm\ we used in the figure the
\chandra\ derived flux estimate. Within the \chandra\ FOV there is one
additional known member with solar mass: Id 790 in the work by DLM94
(easily identifiable in the CMD in Fig.\,\ref{fig:colmag}). The upper
limit on the X-ray luminosity of this star is also given in the
figure.

In the figure, the crosses give the median value of the distributions
(including upper limits) and the size of the vertical bars are
determined by the 25\% and 75\% quantiles.  In Sect.\,\ref{sec:res} we
showed that we expect the optical sample by DLM94 to be complete
within the \chandra\ FOV for $M > 0.8~M_{\sun}$, thus the sample of
the 5 \chandra\ detected solar-mass cluster members plus the upper
limit for the one undetected source is very likely to include all the
cluster members with masses in the $0.8-1.2~M_{\sun}$ range, in this
field.  The median luminosity value for this sample is $1.3 \times
10^{28}$\es; the quantiles bars also refer to this sample.  Including
the two candidate members with $M \sim 0.8 M_{\sun}$ (Ids 136 and 246
-- see Table\,\ref{tab:lx}) does not change the median value;
including the one \xmm\ detected (solar-mass) member outside the
\chandra\ FOV does not change this value either, although it increases
the size of the 75\% quantile bar.  The error on the distance of the
cluster ($430\pm 20$ pc) is relatively small and does not impact
significantly on the position of the median luminosity of NGC~752 in
the plot; for the distance range of $410-450$ pc, the value of the
median X-ray luminosity stays within the two quantile vertical bars.

The X-ray luminosities of the seven points in NGC~752 have a spread of
more than one order of magnitude, confirming the large spread in
luminosity of stars with the same age and similar mass already
observed in the Pleiades, the Hyades (e.g. \citealp{ssk95};
\citealp{msk+96}) and in the sample by \cite{nb98} studied by
\cite{micela02}. The observed spread in $L_{\rm X}$ at a fixed age is
associated to the spread in rotational periods (\citealp{pmm+03}),
which appears to depend on the coupling between the circumstellar disc
and the star in the pre-main sequence phase; in some objects this
coupling prevents a young star from spinning up during its PMS
contraction, yielding a large spread in the initial angular momentum
distribution (e.g. \citealp{bouvier94}; \citealp{ch96}). The analysis
of $\sim$ 500 pre-main sequence and recently arrived main-sequence
stars by \cite{hm05} supports this view.

The median X-ray luminosity of 1.3 $ \times 10^{28}$\es for solar-type
stars in NGC~752 is in good agreement with the decaying trend of X-ray
luminosity from the Hyades to the field stars. The value is about 6
times lower than the median value for the Hyades and 6.5 times
higher than the median value from field stars, consistent with a
steepening of the X-ray luminosity scaling law after the age of the
Hyades.

In a study of nine solar-like G-stars with ages ranging from 70 Myr to
9 Gyr, \cite{ggs97} found $L_{\rm X} \propto t^{-\beta}$ with $\beta
\sim 1.5$, for stars with ages beyond a few 100 Myr, in agreement with
the earlier results by \cite{msv+87}. The study of a sample of 11
late-type stars in the Chandra Deep Field-North by \cite{fhm+04} is
also consistent with this scaling law, for $ 1 < t < 11$ Gyr, although
an excellent fit to their data is found for $\beta=2.0$. \cite{pp04}
find evidence for a very steep decay of the chromospheric activity
between 0.5 and 2 Gyr, their sample at $\sim 2$ Gyr consisting of
seven stars.

Comparison of the median X-ray luminosity of NGC~752 with that of the
Hyades is fully consistent with a scaling law with $\beta \sim
1.5$. This scaling law is also consistent with the decay in X-ray
luminosity from NGC~752 to the field stars, given the age
uncertainties of the field stars.  Since $L_{\rm X} \approx v_{\rm
  rot}^{2}$ (e.g. \citealp{pgr+81}; \citealp{pmm+03}), this result
implies $\alpha \sim 0.75$ for the scaling law of rotational
velocities, for $t \ga 1$ Gyr. This is significantly steeper than
found by \cite{skuma72} and would require a nearly-dipolar magnetic
configuration to be explained in terms of magnetic breaking
(\citealp{kawaler88}). The effect of differential rotation in the
stellar interior and the onset of magnetic saturation, however, may
also play a role (\citealp{kpb+97}) and it is possible that the
coupling efficiency between outer and inner layers of stars weakens
with age or that it is mass dependent (\citealp{barnes03}).

Comparison of rotational velocities in the Pleiades with those in
older clusters such as M34 and the Hyades shows that, within the age
interval of the Pleiades and Hyades, a star's rotational rate
typically decreases less steeply than predicted by a pure magnetic
braking law, that is $\alpha < 0.3$ (\citealp{qam+98}). The decay of
the median $L_{\rm X}$ of stars with mass $M=0.8-1.2~M_{\sun}$ from
the Pleiades to the Hyades reported by \cite{micela02} (the points
shown in Fig.\,\ref{fig:lxvage}) and that reported by \cite{pf05} for
stars with $M=0.9-1.2~M_{\sun}$ for the same clusters, confirm this
result ($\alpha < 0.25$).

As discussed by \cite{qam+98}, a value of $\alpha < 0.3$ could be an
indication that 
angular momentum tapped in the radiative core of slow rotators on the
zero age main sequence resurfaces into the convective envelope between
the Pleiades and Hyades ages.  We know from helioseismology that there is
no gradient between the angular velocities of the core and the
envelope in the Sun, thus our data suggest a change in rotation
regimes of the stellar interior at $t\sim $ 1 Gyr.

The shape of the temporal evolution of the X-ray luminosity of
solar-mass stars also has an important consequence in the evolution of
close-in exoplanets $-$ within 0.5 AU. As shown by \cite{pml08}, the
later the timing of the transition between the two scaling laws of
$L_{\rm X}$, the smaller the fraction of gaseous planets which at 4.5 Gyr
retain most of their initial mass. Our result indicates this
transition to be at around 1 Gyr.

\begin{acknowledgements}

We gratefully acknowledge our referee, Eric Feigelson, for his useful
suggestions which have significantly improved the quality of the
paper. We thank Marco Miceli for providing us the spectral analysis of
the extended source in the \xmm\ FOV, Guido de Marchi for useful
discussions and Andrew Thean for his help with the English style. GM and IP
acknowledge the financial contribution from contract ASI-INAF
I/088/96/0.  This research makes use of data products from the Two
Micron All Sky Survey, which is a joint project of the University of
Massachusetts and the Infrared Processing and Analysis
Center/California Institute of Technology, funded by the National
Aeronautics and Space Administration and the National Science
Foundation.

\end{acknowledgements}


\end{document}